\documentclass[aps,prb,twocolumn,superscriptaddress,amsmath,showpacs,amssymb]{revtex4} 

\usepackage{graphicx}
\usepackage{amsthm,amsmath}
\usepackage{color}

\newcommand{\be}{\begin{eqnarray}}
\newcommand{\ee}{\end{eqnarray}}

\newcommand{\bo}{\boldsymbol}
\newcommand{\lb}{\label}

\newcounter{ichi}
\newcounter{ni}
\newcounter{san}
\newcounter{yon}
\newcounter{go}
\newcounter{roku}
\newcounter{nana}
\newcounter{hati}
\newcounter{kyu}
\begin{document}

%\preprint{APS/123-QED}
\title{Massless collective excitations in frustrated multi-band superconductors}
\author{Keita Kobayashi}
\affiliation{CCSE, Japan Atomic Energy Agency, 5-1-5 Kashiwanoha, Kashiwa, Chiba 277-8587, Japan}
%\author{Masahiko Okumura}
%\affiliation{CCSE, Japan Atomic Energy Agency, 5-1-5 Kashiwanoha, Kashiwa, Chiba 277-8587, Japan}
%\affiliation{Computational Condensed Matter Physics Laboratory, RIKEN, Wako, Saitama 351-0198, Japan}
%\affiliation{Computational Materials Science Research Team, RIKEN AICS, Kobe, Hyogo 650-0047, Japan}
\author{Masahiko Machida}
\affiliation{CCSE, Japan Atomic Energy Agency, 5-1-5 Kashiwanoha, Kashiwa, Chiba 277-8587, Japan}
\affiliation{Computational Materials Science Research Team, RIKEN AICS,
Kobe, Hyogo 650-0047, Japan}
\author{Yukihiro Ota}
\affiliation{CEMS, RIKEN, Saitama 351-0198, JAPAN}
\author{Franco Nori}
\affiliation{CEMS, RIKEN, Saitama 351-0198, JAPAN}
\affiliation{Physics Department, University of Michigan, 
Ann Arbor, Michigan 48109-1040, USA}

\date{\today}% It is always \today, today,
             %  but any date may be explicitly specified

\begin{abstract} 
We study collective excitations in three- and four-band superconductors
 with inter-band frustration, which causes neither $0$ nor $\pi$ inter-band 
phases in the superconducting state. 
%Using a low-energy spin-Hamiltonian originating from a multi-band 
%tight-binding model, we find that a massless Leggett mode 
% occurs in a wide parameter region of this four-band system. 
%This massless mode is related to the fact that the mean-field energy does 
%not depend on a continuous superconducting phase. 
Using a low-energy spin-Hamiltonian originating from a multi-band 
tight-binding model, we find that mass reduction of a Leggett mode
 occurs in a wide parameter region of this four-band system. 
As a limitting case, we have a massless Leggett mode. 
This massless mode is related to the fact that the mean-field energy does 
not depend on a relative phase of superconducting order parameters. 
In other words, we find a link of the massless mode with a degeneracy 
between a time-reversal-symmetry-breaking state (neither $0$ nor $\pi$ phases) 
and a time-reversal-symmetric state (either $0$ or $\pi$ phases).  
Therefore, the mass of collective modes characterizes well the
 time-reversal symmetry in frustrated multi-band superconductors. 
\end{abstract}

% PACS, the Physics and Astronomy
                             % Classification Scheme.
%\keywords{Suggested keywords}%Use showkeys class option if keyword
                              %display desired
\maketitle

%%%%%%%%%%%%%%%%%%%%%%%%%%%%%%%%%%%%
\section{Introduction}
Frustration leads to intriguing phenomena in different physical
systems~\cite{Diep,PI-junction}. 
Multi-band superconductors/superfluids, such as iron-based
materials~\cite{iron,Paglione;Greene:2010} and multi-component
ultra-cold atomic gases~\cite{p-band,Yb,p-band2}, 
can be frustrated many-body systems. 
Frustration in these systems originates from competitive interaction 
between different bands/components, not different spatial sites.  
This curious {\it inter-band frustration} allows a 
time-reversal-symmetry breaking (TRSB) superconducting
state~\cite{TRSB1,TRSB2,TRSB2.5}. 

Collective excitations characterize well an ordered state in many-body
quantum systems. 
The Leggett mode~\cite{Leggett,ota,Carlsterom;Babaev:2011,shi,Stanev:2012,Weston_Babaev,Yanagisawa;Hase:2013} is a characteristic
 collective excitation in multi-band superconductivity, as well as the
 Nambu-Goldstone (NG) mode associated with ${\rm U}(1)$-symmetry breaking,
 and has been studied in multi-band systems such as magnesium 
diboride~\cite{Blumberg;Karpinski:2007}, iron-based
materials~\cite{Burnell;Bernevig:2010}, and atomic gases on a honeycomb
optical lattice~\cite{Zhao;Paramekanti:2006,Tsuchiya;Paramekanti:2012}. 
The mass of the Leggett mode strongly depends on inter-band
couplings~\cite{ota}. 
A recent striking result~\cite{shi} is that the mass in a three-band
system vanishes at the boundary between a time-reversal symmetric
(TRS) state and a TRSB state, changing the strength of the inter-band 
coupling. 

\begin{figure}[t]
\centering
\includegraphics[width=0.85\linewidth]{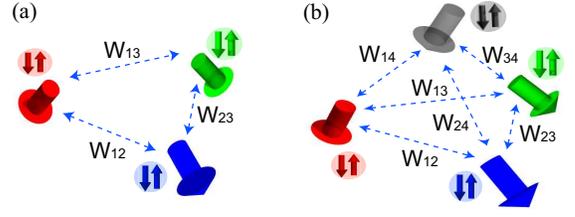}
\caption{(Color online) Schematic diagrams of inter-band configurations
 in (a) three- and (b) four-band systems. 
Each large arrow indicates the orientation of a pseudo-spin formed by a
 fermion-pair particle on a band. 
Since inter-band couplings $W_{\alpha\alpha^{\prime}}$ are repulsive,
 an anti-ferromagnetic interaction occurs between the pseudo-spins. 
See Eq.~(\ref{eq:spin}). }
\label{systematic}
\end{figure}

In this paper, we study the connection between inter-band frustration and
the mass of collective excitations. 
To study properties depending on the number of bands, we
focus on two cases, as seen in Fig.~\ref{systematic}. 
First, we examine a three-band system as a minimal model for showing the
inter-band frustration. 
Second, we study a four-band system as an example which 
shows a feature different from the three-band system. 
Our approach is to make a map from a multi-band tight-binding model to
an effective frustrated spin-Hamiltonian. 
An analogy with a classical spin system is useful for examining 
multi-band superconductors~\cite{TRSB2.5}. 

A mean-field theory of the effective spin-Hamiltonian allows us to
calculate the superconducting-phase configurations and the collective
excitations. 
Varying the strength of the inter-band couplings, we obtain a phase 
diagram of the superconducting state. 
The massless Leggett mode is found at the phase boundaries between the
TRSB and TRS states. 
This result is consistent with the result by Lin and Hu~\cite{shi}. 
The main result in this paper is that in the four-band system a
massless Leggett mode occurs in a parameter region other than the
TRSB-TRS phase boundaries.  
In this region, the mean-field energy for a TRSB state is equal to the
one for a TRS state. 
Therefore, this massless behavior is related to the degenerate
superconducting states. 
Moreover, we characterize this massless mode, from the viewpoint of
inter-band symmetry. 
Thus, we claim that the mass of collective excitations gives an insight
into spontaneous-symmetry breaking in the presence of inter-band
frustration. 

This paper is organized as follows. 
The effective spin Hamiltonian is derived from a multi-band tight-binding
model in Sec.~\ref{sec:hamiltonian}. 
The formulae for calculating the superconducting order parameter and the
collective excitations are derived, with mean-field approximation. 
In Sec.~\ref{sec:results}, we solve the resultant formulae in a
spatially homogeneous case. 
We show that the massless behaviors of the Leggett
mode are associated with energy degeneracy between the TRSB state and the
TRS state. 
Furthermore, we discuss an effect of quantum fluctuations on the 
massless modes in Sec.~\ref{sec:discussion}.  
Section \ref{sec:summary} is devoted to the summary. 

\section{Effective Hamiltonian with anti-ferromagnetic XY interaction}
\label{sec:hamiltonian}
An effective Hamiltonian is derived from a multi-band tight-binding
model, via the second order perturbation. 
This effective model explicitly shows the presence of inter-band
frustration, in terms of anti-ferromagnetic $XY$ interaction. 
Using the mean-filed approximation, we show the formulae for calculating
the superconducting order parameters and the collective excitations in a
spatially homogeneous case. 
We also define a witness for the TRSB state, scalar chiral order
parameter. 
In the subsequent section, we will calculate these equations numerically. 

The Hamiltonian is 
\be
H
=
\sum_{\alpha} \sum_{\sigma=\uparrow,\downarrow} h_{\alpha,\sigma} 
+
\sum_{\alpha,\alpha^{\prime}} v_{\alpha,\alpha^{\prime}},
\label{eq:tight_b_h}
\ee
with
\be
&&
h_{\alpha,\sigma}
=
-\sum_{< \bo{i},\bo{j} >}
t_{\alpha}c_{\alpha,\sigma,\bo{i}}^\dagger c_{\alpha,\sigma,\bo{j}}
-\sum_{\bo{i}}
\mu \,c_{\alpha,\sigma,\bo{i}}^\dagger c_{\alpha,\sigma,\bo{i}}, \\
&&
v_{\alpha,\alpha^{\prime}}
=
\sum_{\bo{i}}W_{\alpha\alpha^{\prime}}
c_{\alpha,\uparrow,\bo{i}}^\dagger c_{\alpha,\downarrow,\bo{i}}^\dagger
c_{\alpha',\downarrow,\bo{i}}c_{\alpha',\uparrow,\bo{i}}.
\ee
The spatial site is $\bo{i}=(i_{x},i_{y},i_{z})$. 
The electron creation (annihilation) operator is
$c_{\alpha,\sigma,\bo{i}}^\dagger$ 
($c_{\alpha,\sigma,\bo{i}}$) for the $\alpha$th band on $\bo{i}$. 
The hopping matrix element and the chemical potential
are, respectively, $t_{\alpha}$ and $\mu$. 
The intra-band coupling $W_{\alpha\alpha}$ is negative (attractive
interaction), while the inter-band coupling $W_{\alpha\alpha^{\prime}}$
($\alpha\neq \alpha^{\prime}$) is positive (repulsive interaction). 

Our approach for deriving an effective model from
Eq.~(\ref{eq:tight_b_h}) is the second-order Brillioun-Wigner
perturbation. 
Since strong intra-band coupling produces condensates, our choice of a
free Hamiltonian is 
\(
H_{0} = \sum_{\alpha} v_{\alpha,\alpha}
\). 
The attractive-repulsive transformation~\cite{ARtrans} makes
Eq.~(\ref{eq:tight_b_h}) a half-filled system. 
This transformation is defined by 
\mbox{
\(
c_{\alpha,\uparrow,\bo{i}}
=
\bar{c}_{\alpha,\uparrow,\bo{i}}
\)} 
and 
\mbox{
\(
c_{\alpha,\downarrow,\bo{i}}
=
\exp(-i\, \bo{q}\cdot \bo{x}_{\bo{i}})\,
\bar{c}_{\alpha,\downarrow,\bo{i}}^\dagger
\)}, with a reciprocal vector $\bo{q}$ satisfying 
\(
\exp[
i\, \bo{q}\cdot (\bo{x}_{\bo{i}+\bo{1}_{l}} - \bo{x}_{\bo{i}})
]
=-1
\), for $l=x,y,z$, where 
\(
\bo{x}_{\bo{i}} = \sum_{l}i_{l}\bo{a}_{l}
\) 
and 
\(
\bo{1}_{l} = \bo{a}_{l}/|\bo{a}_{l}|
\). 
The lattice vector along $l$-axis is $\bo{a}_{l}$. 
The ground-state subspace of $H_{0}$ is 
\(
\mathcal{H}_{{\rm g}}=\otimes_{\alpha,\bo{i}}
\{
\bar{c}_{\alpha,\uparrow,\bo{i}}^{\dagger}|\bar{0}\rangle, 
\,
\bar{c}_{\alpha,\downarrow,\bo{i}}^{\dagger}|\bar{0}\rangle
\}
\), 
and the excited-state subspace is 
\(
\mathcal{H}_{\rm e}=\otimes_{\alpha,\bo{i}}
\{
|\bar{0}\rangle ,\,
\bar{c}_{\alpha,\downarrow,\bo{i}}^{\dagger}
\bar{c}_{\alpha,\uparrow,\bo{i}}^{\dagger}
|\bar{0}\rangle 
\}
\). 
The ket vector $|\bar{0}\rangle$ is defined by 
\(
\bar{c}_{\alpha,\sigma,\bo{i}}|\bar{0}\rangle =0
\). 
The effective Hamiltonian is 
\(
H_{\rm eff}
=
PVP - (PVQ)H_{0}^{-1}(QVP)
\), 
with $V=H-H_{0}$. 
The projector onto $\mathcal{H}_{\rm g}$ ($\mathcal{H}_{\rm e}$) is $P$
($Q$). 

Let us write $H_{\rm eff}$ in terms of the pseudo-spin $1/2$
operators defined by
\(
\bar{S}_{\alpha,\bo{i}}^{(+)}
=
\bar{c}_{\alpha,\uparrow,\bo{i}}^{\dagger}
\bar{c}_{\alpha,\downarrow,\bo{i}}
\)
\,, 
\(
\bar{S}_{\alpha,\bo{i}}^{(-)}
=
[\bar{S}_{\alpha,\bo{i}}^{(+)}]^\dagger
\), 
and 
\(
\bar{S}_{\alpha,\bo{i}}^{(z)}
=
(
\bar{c}_{\alpha,\uparrow,\bo{i}}^{\dagger}
\bar{c}_{\alpha,\uparrow,\bo{i}}
-
\bar{c}_{\alpha,\downarrow,\bo{i}}^{\dagger}
\bar{c}_{\alpha,\downarrow,\bo{i}}
)/2
\). 
These operators represent a fermion-pair particle. 
The 2nd perturbation term leads to the Heisenberg Hamiltonian
with exchange interaction 
\mbox{
\(
J_{\alpha}= 2t_\alpha^{2} /|W_{\alpha\alpha}|
\)}. 
The contribution from $W_{\alpha\alpha^{\prime}}$ 
($\alpha\neq \alpha^{\prime}$) appears as the 1st perturbation
term since  
\mbox{
\(
Pv_{\alpha,\alpha^{\prime}}Q =0
\)} for $\alpha\neq \alpha^{\prime}$. 
Thus, 
\be
H_{\rm eff}
&=&
\sum_\alpha \sum_{<\bo{i},\bo{j}>}
J_{\alpha} \left[
\bar{S}_{\alpha,\bo{i}}^{(z)}\bar{S}_{\alpha,\bo{j}}^{(z)}
+
\bar{S}_{\alpha,\bo{i}}^{(+)}\bar{S}_{\alpha,\bo{j}}^{(-)}
\right] 
\nonumber \\
&&
+\sum_{\alpha \neq\alpha'}\sum_{\bo{i}}
W_{\alpha\alpha'}
\bar{S}_{\alpha,\bo{i}}^{(+)}\bar{S}_{\alpha',\bo{i}}^{(-)}
-\sum_{\alpha,\bo{i}}2\bar{\mu}_{\alpha}\bar{S}_{\alpha,\bo{i}}^{(z)}, 
\lb{eq:spin}
\ee
with \mbox{$\bar{\mu}_{\alpha}=\mu+|W_{\alpha\alpha}|/2$}. 
The inter-band interaction is regarded as an anti-ferromagnetic
$XY$-interaction. 

We examine Eq.~(\ref{eq:spin}), using the mean-field approach with
spatial uniformity. 
Let us rewrite the pseudo-spin-$1/2$ operators, in terms of
$b_{\alpha,\bo{i}}$, such that 
\mbox{
\(
\bar{S}_{\alpha,\bo{i}}^{(+)}
=\exp(-{\rm i}\bo{q}\cdot \bo{x}_{\bo{i}}) b_{\alpha,\bo{i}}
\)}
and 
\mbox{
\(
\bar{S}_{\alpha,\bo{i}}^{(z)}
=(1/2)-b_{\alpha,\bo{i}}^\dagger b_{\alpha,\bo{i}}
\)}. 
We find that 
\mbox{
\(
[ b_{\alpha,\bo{i}},b_{\alpha^{\prime},\bo{i}^{\prime}}^\dagger]
=
(1-b_{\alpha,\bo{i}}^\dagger b_{\alpha,\bo{i}})
\delta_{\bo{i}\bo{i}^{\prime}}\delta_{\alpha\alpha^{\prime}}
\)}
and 
\(
b_{\alpha,\bo{i}}^2=0
\). 
In the dilute limit $\langle b_{\alpha,\bo{i}}^\dagger
b_{\alpha,\bo{i}}\rangle \ll 1$\,, $b_{\alpha,\bo{i}}$ can be
regarded as a standard bosonic operator.  
Using 
\(
\langle b_{\alpha,\bo{i}}\rangle=\Delta_{\alpha,\bo{i}}
\), 
we obtain the mean-field energy $E_{\rm c}$ as a function of
$\Delta_{\alpha,\bo{i}}$. 
For the uniform order parameters
($\Delta_{\alpha,\bo{i}}=\Delta_{\alpha}$),  
$\Delta_{\alpha}$ is determined by 
\(
(\partial E_{\rm c}/\partial \Delta_{\alpha}^{\ast})=0
\), namely, 
\be
-2J_{\alpha}D(1-2|\Delta_{\alpha}|^2)\Delta_{\alpha}
+\sum_{\alpha' \neq\alpha }W_{\alpha\alpha'}\Delta_{\alpha'}-
\nu_\alpha \Delta_{\alpha}=0, 
\label{eq:GP} 
\ee
where $\nu_\alpha=2DJ_{\alpha}-2\bar{\mu}_{\alpha}$ and $D$ is the
dimension of the system. 
The collective excitations for momentum $\bo{k}$ are calculated,
combining the resultant gaps with the Bogolubov de-Gennes equation 
\be 
T_{\bo{k}}\bo{Y}_{\bo{k}}=\omega_{\bo{k}}\bo{Y}_{\bo{k}}\,,
\label{eq:BdG}
\ee
with 
\(
T_{\bo{k}} = 
\tau_{z}\otimes \mathcal{L} 
+ \tau_{x}\otimes i {\rm Im}\,\mathcal{M} 
+ \tau_{y}\otimes i {\rm Re}\,\mathcal{M}
\). 
$\bo{Y}_{\bo{k}}$ is a $2N$-complex vector, where
$N$ is the number of bands. 
The $2\times 2$ Pauli matrices ($\tau_{x},\tau_{y},\tau_{z}$) 
represent the so-called particle-hole
symmetry of the Bogoliubov-de Gennes equation.
The $N\times N$ matrices $\mathcal{L}$ and $\mathcal{M}$ are defined as, 
\be 
\mathcal{L}_{\alpha\alpha'}
&=&-2\delta_{\alpha,\alpha'}
\sum_l 
[
\varepsilon_{\alpha,k_l}-2(\varepsilon_{\alpha,k_l}
-\varepsilon_{\alpha,0})|\Delta_{\alpha}|^2
]
\nonumber \\
&&-\delta_{\alpha,\alpha'}\nu_\alpha
+(1-\delta_{\alpha,\alpha'})W_{\alpha\alpha'}\,, \\
\mathcal{M}_{\alpha\alpha'}
&=&\delta_{\alpha,\alpha'}
\sum_l4\varepsilon_{\alpha,k_l} \Delta_\alpha^2.
\ee
The coefficient $\varepsilon_{\alpha,k_{l}}$ is the Fourier-transformed
hopping matrix element, 
\(
\varepsilon_{\alpha,k_l}=J_\alpha \cos(k_la_l)
\), 
with lattice constant $a_{l}(=|\bo{a}_{l}|)$. 

The superconducting states are classified by the scalar chiral order
parameter\,\cite{Wen;Zee:1989,note1}
\begin{equation}
 \chi = \sum_{\alpha_{1}<\alpha_{2}<\alpha_{3}}
  |\left\langle
\bar{\bo{S}}_{\alpha_1}\cdot
(\bar{\bo{S}}_{\alpha_2}\times\bar{\bo{S}}_{\alpha_3})
\right\rangle |.
\label{eq:schiral}
\end{equation}
Under the mean-field approximation and the dilute limit, the components
of the pseudo-spin $1/2$ vector $\bar{\bo{S}}_{\alpha}$ are 
\(
\bar{\bo{S}}_{\alpha} 
\simeq 
\langle 
\bar{\bo{S}}_{\alpha} 
\rangle
=
\,^{\rm t}(
\Delta_{\alpha}^{\rm R},
\Delta_{\alpha}^{\rm I},
1/2-|\Delta_{\alpha}|^2
)\simeq
\,^{\rm t}(
\Delta_{\alpha}^{\rm R},
\Delta_{\alpha}^{\rm I},
1/2
)
\), 
where $\Delta_{\alpha}^{\rm R}$ and $\Delta_{\alpha}^{\rm I}$ are,
respectively, the real and the imaginary parts of $\Delta_{\alpha}$. 
The inter-band phases (e.g., 
\(
\Delta_{1}^{\rm R}\Delta_{2}^{\rm I}
-
\Delta_{1}^{\rm I}\Delta_{2}^{\rm R}
\)) are important for determining the TRSB state. 
We sum up such quantities over all the band indices in 
Eq.~(\ref{eq:schiral}).
We note that $\chi=0$ when $\Delta_{\alpha}^{\rm I}=0$ for all
$\alpha$. 

\begin{figure}[t]
\centering
\includegraphics[width=1.0\linewidth]{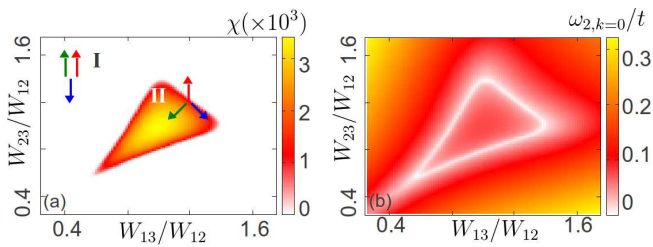}
\caption{(Color online) Density profiles of (a) a scalar chiral order parameter and (b)
 Leggett-mode mass, in a three-band superconductor, varying
 inter-band couplings $W_{13}/W_{12}$ and $W_{23}/W_{12}$.  
In (a), the arrows show typical superconducting-phase configurations.}
\label{3-band}
\end{figure}

\section{Mass reduction of a Leggett mode by inter-band frustration}
\label{sec:results}
We calculate the scalar chiral order parameter on a 2D square lattice
($D=2$ and $a_{l}=a$), numerically solving Eq.~(\ref{eq:GP}), 
according to the imaginary-time evolution method~\cite{ITE,note2}. 
We also evaluate the collective modes by direct diagonalization of
Eq.~(\ref{eq:BdG}). 
Here, we consider in a highly symmetric case $t_{\alpha}=t$ for
simplicity and 
focus on a strong intra-band interaction case,
$W_{\alpha\alpha}/t=-6$, to ensure the validity of Eq.~(\ref{eq:spin}). 
Throughout this paper, we set $J_{\alpha}/t= J/t =1/3$. 
The condensate particle-density is also fixed as 
$n_{\rm c}=\sum_\alpha|\Delta_{\alpha}|^2=0.1$. 
The number of the collective modes depends on $N$. 
We will denote the NG mode as $\omega_{1,\bo{k}}$. 
The others correspond to the Leggett modes. 

First, we show the results for the three-band case. 
Figure \ref{3-band}(a) shows the presence of different parameter
regions. 
In region $\mbox{I}$ the TRS states occur 
($\chi =0$), whereas in region $\mbox{I\!I}$ the TRSB states occur
($\chi \neq 0$). 
In region $\mbox{I}$ a sign change (anti-parallel arrangement of
pseudo-spins) occurs between the gaps. 
In region $\mbox{I\!I}$ a typical phase-configuration is that each
relative superconducting phase is $2\pi/3$. 
In other words, each pseudo-spin directs from the
center to the vertex of an equilateral triangle. 
Figure \ref{3-band}(b) shows that the mass of the Leggett mode
($\omega_{2,\bo{k}=0}$) vanishes at the TRSB-TRS phase boundaries. 
These results are consistent with the results of a weak-coupling
model\,\cite{shi}. 
The phase transition between the TRS and the TRSB state is the
2nd-order one, as shown by Lin and Hu~\cite{shi}. 
The fluctuation developed at the critical point may lead to this massless
behavior. 

\begin{figure}[t]
\centering
\includegraphics[width=1.0\linewidth]{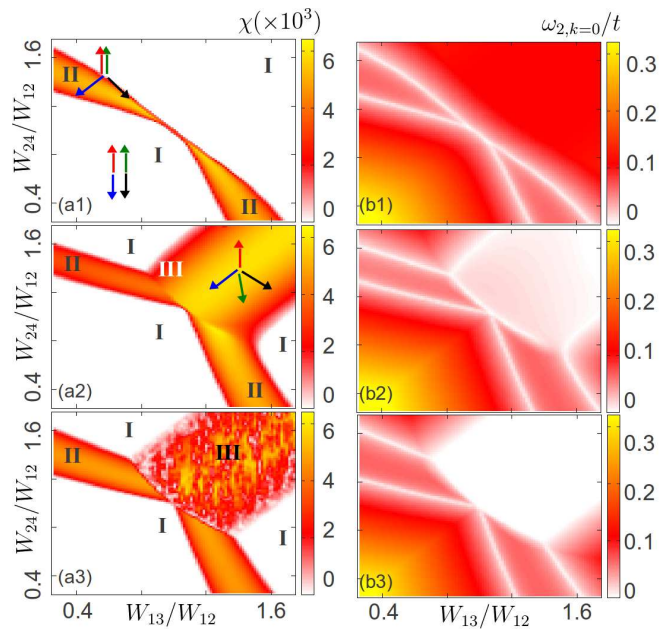}
\caption{(Color online) Density profiles of (a1,a2,a3) a scalar chiral order paramter
 and (b1,b2,b3) Leggett-mode mass, in a four-band superconductor, varying
 inter-band couplings $W_{13}$ and $W_{24}$. 
 The other inter-band couplings are fixed as
 $W_{12}=W_{34}=0.23$\,, $W_{23}=W_{14}=0.2$ in (a1,b1), 
 $W_{12}=W_{23}=0.23$\,, $W_{34}=W_{14}=0.2$ in (a2,b2), 
 and $W_{12}=W_{23}=W_{34}=W_{14}=0.2$ in (a3,b3). The arrows in (a1,a2)
 show typical superconducting-phase configurations like
 Fig.\,\ref{3-band}.}
\label{4-band}
\end{figure}

\begin{figure}[t]
\centering
\includegraphics[width=1.0\linewidth]{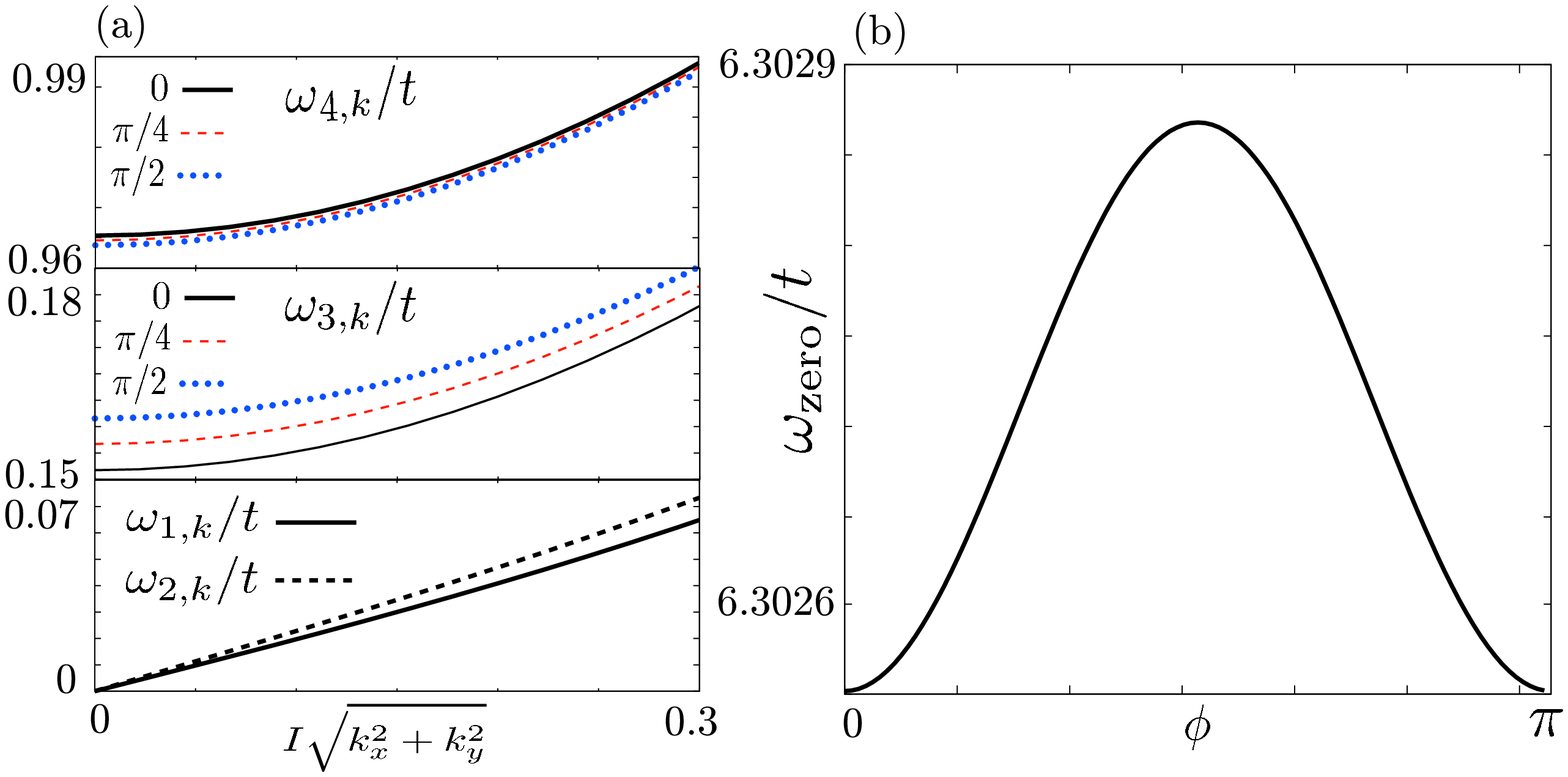}
\caption{(Color online) (a) Dispersion relations for collective excitations, with
 relative phases $\phi=0,\pi/4,\pi/2$. $\omega_{1,\bo{k}}$ and
 $\omega_{2,\bo{k}}$ are independent of $\phi$. (b) Zero-point energy of
 the collective excitations, varying $\phi$. In both figures, inter-band
 couplings are set as $W_{13}/W_{12}=1.2$, $W_{24}/W_{12}=1.3$, and
 $W_{12}=W_{23}=W_{34}=W_{14}=0.2$. }
\end{figure} 

Now, let us show the four-band case. 
We change $W_{13}$ and $W_{24}$, with fixed 
$W_{12},W_{23},W_{34}$ and $W_{14}$. 
From the viewpoint of Fig.~\ref{systematic}, the length of four sides in a
tetrahedron is fixed. 
First, we show similar features to the three-band case. 
Figure \,\ref{4-band}(a1) shows the results for $W_{12}=W_{34}=0.23$
and $W_{23}=W_{14}=0.2$. 
The TRSB state appears in region $\mbox{I\!I}$, whereas the TRS
states occur in the other regions. 
The phase configuration in region $\mbox{I\!I}$ is similar to the
three-band case, although two of the pseudo-spins are aligned ($0$-phase
shift). 
We also find that the mass of the Leggett mode vanishes at the TRSB-TRS
phase boundaries, as seen in Fig.~\ref{4-band}(b1). 
Changing the condition for the fixed inter-band couplings, different
features appear. 
Let us consider the case of $W_{12}=W_{23}=0.23$ and $W_{34}=W_{14}=0.2$. 
Figure \ref{4-band}(a2) shows the presence of a curious area (region 
${\rm I\!I\!I}$), where the time-reversal symmetry is fully broken. 
In other words, every relative phase is neither $0$ nor $\pi$. 
Figure \ref{4-band}(b2) shows that the mass of the Leggett mode is
close to zero inside this region. 
We can also find that $\omega_{2,\bo{k}=0}$ and $\omega_{3,\bo{k}=0}$
become zero near the phase boundaries between 
$\mbox{I\!I}$ and ${\rm I\!I\!I}$ (no figure shown for $\omega_{3,\bo{k}}$). 
A more exotic feature appears in the identical inter-band interaction
$W_{12}=W_{23}=W_{34}=W_{14}=0.2$. 
Figure \ref{4-band}(a3) shows that $\chi$ randomly changes in
region $\mbox{I\!I\!I}$\cite{note2}. 
$\omega_{2,\bo{k}}$ is massless in this wide area,
not restricted near the phase boundaries. 

We examine region $\mbox{I\!I\!I}$ in Fig.~\ref{4-band}(a3) in detail. 
Since $W_{13}>W_{12}$, a $\pi$-shift may occur between $\Delta_{1}$ and
$\Delta_{3}$. 
Similarly, the condition $W_{24}>W_{12}$ means 
$\Delta_{4}=e^{\pm i\pi}\Delta_{2}$. 
Moreover, since $W_{12}=W_{34}$, the relative phase between $\Delta_{1}$ and
$\Delta_{2}$ should be equal to the one between $\Delta_{3}$ and
$\Delta_{4}$. 
Therefore, we construct a solution of Eq.~(\ref{eq:GP}) in region
${\rm I\!I\!I}$, with ansatz 
\be
&&
\bo{\Delta}
=
(|\Delta_{+}|,e^{i\phi}|\Delta_{-}|, e^{i\pi}|\Delta_{+}|,
e^{i(\phi+\pi)}|\Delta_{-}|). 
\label{eq:exp_form_ordp_4}
\ee
The global phase is fixed so that $\Delta_{1}$ is real. 
Substituting this expression into Eq.~(\ref{eq:GP}), we find that 
\mbox{
\(
|\Delta_{\pm}|
=
(1/2)\sqrt{n_{\rm c} \pm(W_{13}-W_{24})/2JD}
\)}, but the relative phase is not fixed. 
This result indicates that the mean-field energy for
Eq.~(\ref{eq:exp_form_ordp_4}) is independent of the continuous
parameter $\phi$, and a degeneracy exists between the TRSB and the TRS
states. 
Thus, the massless behavior in region $\mbox{I\!I\!I}$ is related to
a degeneracy.  
The occurrence of such an exotic massless
mode and a degeneracy between ground states were pointed out by several
authors~\cite{Song,Uchino,Cai,Castro}.

A symmetry analysis of Eq.~(\ref{eq:BdG}) leads to insights into the
massless Leggett mode.  
The identical inter-band couplings ($W_{12}=W_{23}=W_{34}=W_{14}$) and
the order parameters (\ref{eq:exp_form_ordp_4}) indicate the 
presence of a symmetric property in Eq.~(\ref{eq:BdG}). 
We find that 
\(
\mathcal{L}
=
\openone \otimes \mathcal{L}_{0} + \eta_{x}\otimes \mathcal{L}_{x}
\)
and 
\(
\mathcal{M}
=
\openone \otimes \mathcal{M}_{0}
\), 
with the $x$-component of the $2 \times 2$ Pauli matrices, 
$\eta_{x}$ and complex $2\times 2$ matrices $\mathcal{L}_{0}$,
$\mathcal{L}_{x}$, and $\mathcal{M}_{0}$. 
Hence, $\eta_{x}$ commutes with $T_{\bo{k}}$.  
We mention that $\eta_{x}$ corresponds to swap between the upper
2-band and the lower 2-band blocks. 
After a permutation, we rewrite $T_{\bo{k}}$ as 
\begin{equation}
 T_{\bo{k}} 
= 
\frac{\openone+\eta_{x}}{2}\otimes T_{+,\bo{k}}
+
\frac{\openone-\eta_{x}}{2}\otimes T_{-,\bo{k}}, \label{eq:BdGop}
\end{equation}
with
\(
T_{\pm,\bo{k}}
=
{\rm tr}_{\eta}
[
 T_{\bo{k}} 
(\openone \pm \eta_{x})
]/2
\). 
The symbol ${\rm tr}_{\eta}$ means the trace over $\eta$-basis. 
The characteristic polynomial of $T_{-,\bo{k}}$ is written by 
$f(z)= \sum_{n=0}^{4} c_{n}(\bo{k}) z^{4-n}$, with $c_{0}=1$. 
Since we can find that the coefficients for $n=1,\,2,\,3$ are 
zero when $\bo{k}=0$, 
$T_{\bo{k}}$ has two zero modes, one of which is the NG  
mode, while the other of which is the massless Leggett mode. 
%The spontaneous-breaking of ${\rm U}(1)$-symmetry indicates that
%either $T_{+,\bo{k}}$ or $T_{-,\bo{k}}$ must have a degenerate zero
%eigenvalue (the Nambu-Goldstone mode). 
%We note that this degeneracy is related to the particle-hole symmetry of the
%Bogoliubov-de Gennes equation. 
%In the present case, we find that $T_{-,\bo{k}}$ has two zero
%modes, since the coefficients in the characteristic polynomial of 
%$T_{-,\bo{k}}$ are zero when $\bo{k}=0$. 
%One of the zero modes is the massless Leggett mode. 
Thus, the present massless Leggett mode belongs to the same subspace as 
the NG mode, and is regarded as a quasi-NG mode. 

\section{Discussion}
\label{sec:discussion}
We refer to an effect of quantum fluctuations on the
ground-state degeneracy. 
The simplest approach to take such corrections is to add the zero-point
energy of the collective excitations to the mean-field energy. 
The correction can be written as 
\(
\omega_{\rm zero}
=
\sum_{\alpha}\sum_{\bo{k}} \omega_{\alpha,\bo{k}}/I
\), with the total number $I$ of the spatial sites. 
Let us examine this correction in region $\mbox{I\!I\!I}$ of
Fig.~\ref{4-band}(a3). 
Figure 4(a) shows that $\omega_{3,\bo{k}}$ and $\omega_{4,\bo{k}}$
depend on $\phi$, whereas the others not so. 
Figure 4(b) shows $\omega_{\rm zero}$ has minimum values at either $0$
or $\pi$. 
In other words, the massive collective modes in region $\mbox{I\!I\!I}$
make a selection of a true ground-state. 
The TRS state is preferable in region ${\rm I\!I\!I}$, owing to
$\omega_{\rm zero}$.

The above consideration indicates that our quasi-NG mode may obtain some
mass originating from quantum fluctuations. 
This point is also discussed in a different system, spinor 
Bose-Einstein condensate~\cite{Uchino}. 
Nevertheless, the mass of the Leggett mode is a good indicator of
inter-band frustration. 
Indeed, our calculations show that the mass of the Leggett mode
drastically reduces (almost zero) when strong competition between the
inter-band couplings, even though the ground-state degeneracy is
absent. 
See region ${\rm I\!I\!I}$ of Fig.~\ref{4-band}(b2), for example. 
In this region, only the TRSB state occurs; it means that strong
inter-band frustration appears. 
Thus, although the Leggett mode does not become a compelte massless mode
in the presence of quantum fluctuations, one may observe a significant
mass reducing behavior of a Leggett mode. 
When the long-range Coulomb interaction exists, the situation becomes
much clearer. 
Typically, the NG mode obtains the mass via the Anderson-Higgs
mechanism; the massive plasma excitations may appear. 
However, since the Leggett modes are related to neutral superfluid-phase
fluctuations~\cite{ota}, one may observe low-energy excitations related
to the Leggett mode with tiny mass, whenever strong inter-band
frustration exists. 
Therefore, we expect that the mass reducing behavior of the Leggett mode
predicted by the present mean-field analysis is robust against quantum
fluctuations and the gauge filed. 
A more systematic study about different fluctuations is an 
interesting future work.

\section{Summary}
\label{sec:summary}
We have examined the collective excitations in three- and
four-band superconductors. 
Using an effective spin Hamiltonian, we showed that inter-band
frustration induces two kinds of massless Leggett modes, and
clarified their physical origin.  
The mass of a collective mode characterizes well the time-reversal
symmetry of frustrated multi-band superconductors. 
%In this paper, we examined the ground-state properties and the
%collective excitations in the three- and four-band superconductors. 
%Using a map from the multi-band tight-binding model with repulsive pair
%hopping to the frustrated spin model, we showed the non-trivial features
%including the occurrence of the TRSB state and the massless Leggett modes.  
%These results could be tested in various multi-band systems, especially
%multi-component ultra-cold fermi gases with high controllability. 
%One of the interesting future works is to construct a theoretical
%approach with quantum corrections. 
%A numerical simulation with the density matrix renormalization
%method\,\cite{DMRG,DMRG:1993} will be useful for the systematic
%examinations of quantum corrections. 
%Also, an analogy with a spin tube and the Heisenberg model on a
%triangle lattice is interesting. 

%{\color{red}{\it Note added.}---
%After the submission of this manuscript, 
%we noticed that similar results were obtained in Ref.\,\cite{Weston_Babaev}.
%These authors have studied collective excitation of frustrated multi-band superconductors 
%by Ginzburg-Landau approach.}

\begin{acknowledgments}
We thank M. Okumura and H. Nakamura for useful discussions. 
This work was partially supported by MEXT Strategic Programs for
Innovative Research, and the Computational Materials Science Initiative,
Japan. 
We are indebted to T. Toyama for his support. 
Y.O. is partially supported by the Special Postdoctoral Researchers
Program, RIKEN. 
F.N. acknowledges partial support from the ARO, RIKEN iTHES project, 
JSPS-RFBR Contract No. 12-02-92100, Grant-in-Aid for Scientific Research
(S), MEXT Kakenhi on Quantum Cybernetics, and Funding Program for
Innovative R\&D on S\&T.
\end{acknowledgments}   
%\appendix

%\newpage %Just because of unusual number of tables stacked at end


\begin{thebibliography}{99}
\bibitem{note2}
We randomly change the initial order parameter, 
	solving Eq.~(\ref{eq:GP}) in the imaginary-time evolution
	method.
Whenever there is no degeneracy between the TRSB state and the TRS state, the
	unique solution is determined for every initial choice. 
However, when the degeneracy exits, the
	resultant scalar chiral order parameter $\chi$ depends on the
	choice, as a result $\chi$ randomly changes, as
	seen in region $\mbox{I\!I\!I}$ of Fig.~\ref{4-band}(a3).
\end{thebibliography}
\end{document}